\documentstyle[11pt,epsf,paspconf]{article}
\input psfig 
\def\wisk#1{\ifmmode{#1}\else{$#1$}\fi}
\def\be{\begin{equation}}
\def\ee{\end{equation}}
\def\bea{\begin{eqnarray}}
\def\eea{\end{eqnarray}}
\def\gtrsim{\mathrel{\hbox{\rlap{\hbox{\lower4pt\hbox{$\sim$}}}\hbox{$>$}}}}
\def\lesssim{\mathrel{\hbox{\rlap{\hbox{\lower4pt\hbox{$\sim$}}}\hbox{$<$}}}}
\def\lsim   {\wisk{_<\atop^{\sim}}}
\def\gsim   {\wisk{_>\atop^{\sim}}}
\def\T{T_{o}}
\def\by{\times}

\def\oo{\Omega_{o}}
\def\ol{\Omega_{\Lambda}}
\def\ob{\Omega_{B}}
\def\oc{\Omega_{CDM}}
\def\oh{\Omega_{HDM}}
\def\on{\Omega_{\nu}}
\def\og{\Omega_{\gamma}}
\def\etal{{\em et al.~}}
\def\apj{{\em Ap. J.}}
\def\mnras{{\em M.N.R.A.S.}}

\def\putboxedtopfigs{
\makebox{
\medskip
\noindent
\parbox[l]{2.6truein}{\epsfxsize=2.6truein\epsfbox{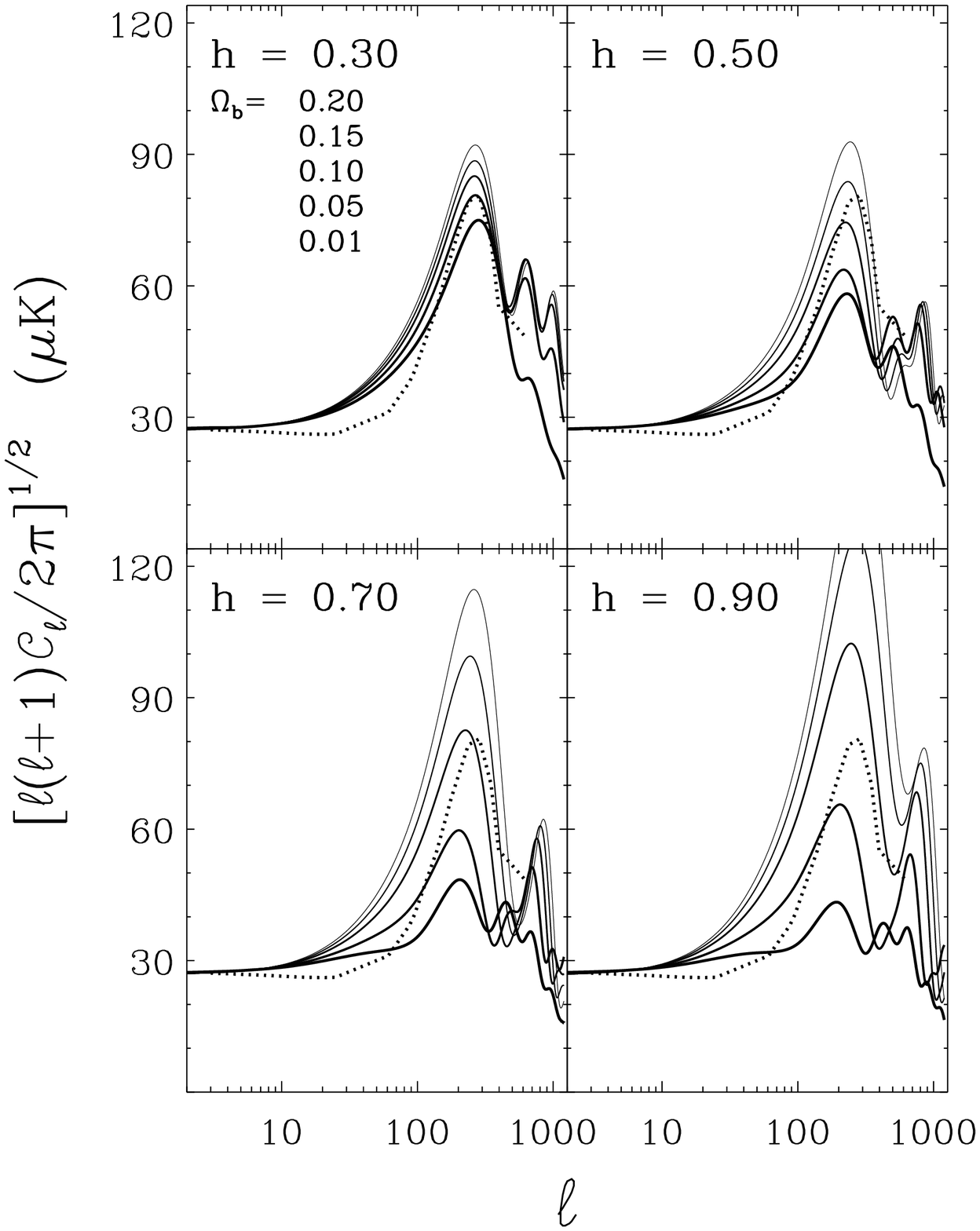}}
\hglue0.1truein
\parbox[r]{2.6truein}{\epsfxsize=2.6truein\epsfbox{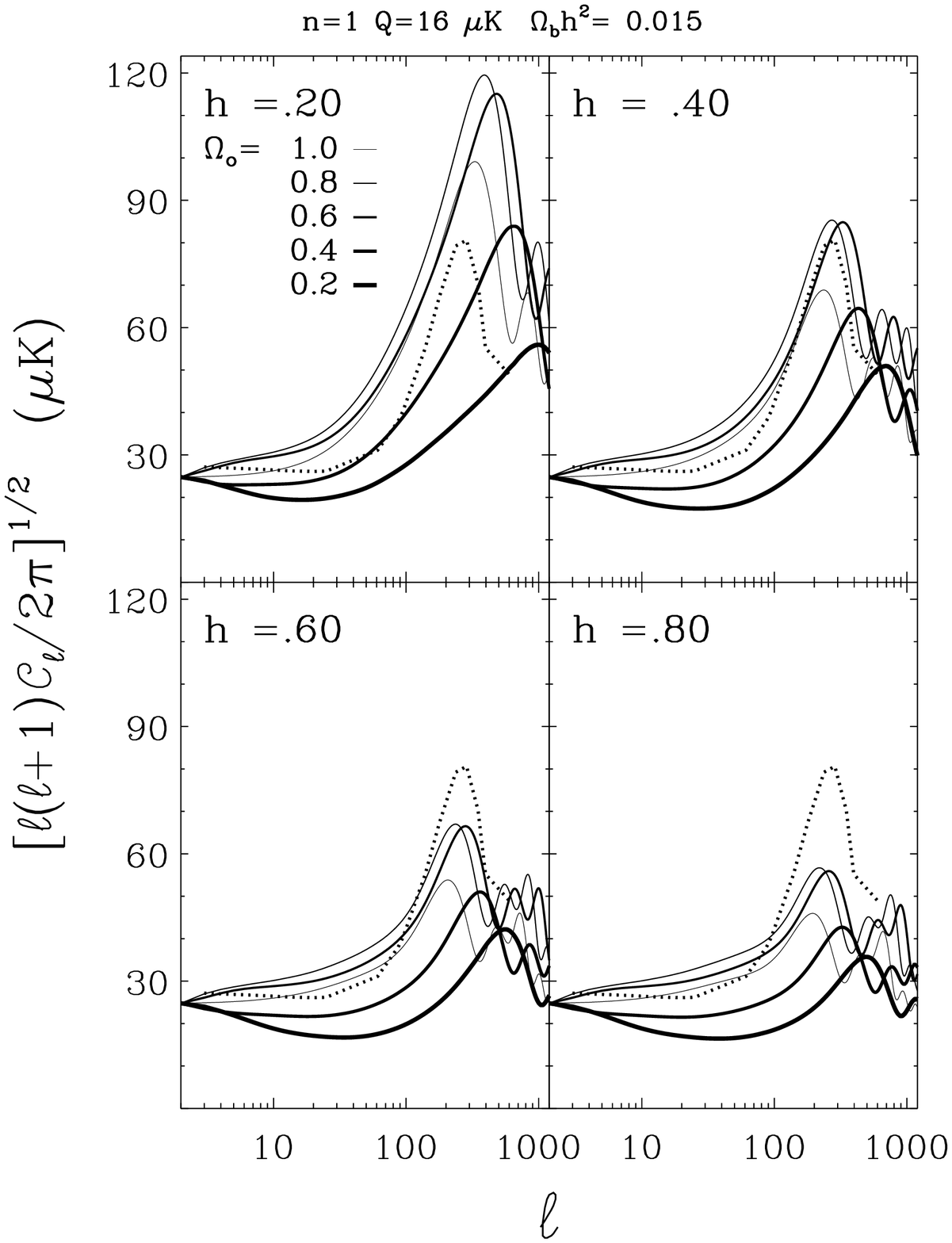}}
\smallskip
\label{fig:dependence}
}
}
\def\putboxedbottomfigs{
\makebox{
\medskip
\noindent
\parbox[l]{2.6truein}{\epsfxsize=2.6truein\epsfbox{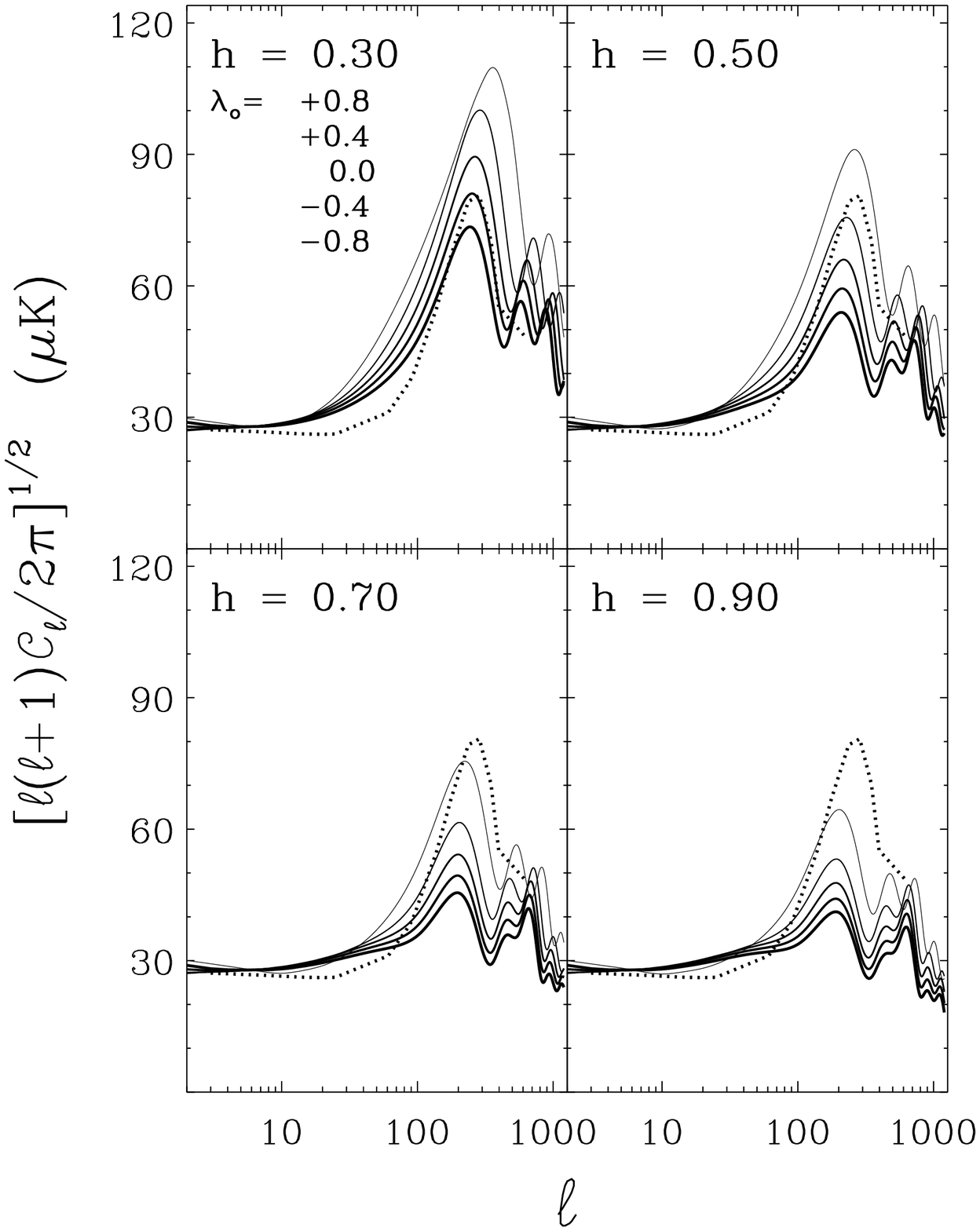}}
\hglue0.05truein
\parbox[r]{2.65truein}{{\footnotesize 
 {\bf Figure 5b. Power Spectra Parameter Dependence}
The dotted line is the polynomial fit to the data in
Figure \protect\ref{fig:data} and is the same in all panels as 
a reference.
Top left: 
in each panel $h$ is fixed while $\ob$ takes on the values indicated.
The largest values of $\ob$ have the largest Doppler peaks.
Notice that as  $h$ increases, the peak amplitude increases for 
large $\ob$ but decreases for small $\ob$; thus at high $h$ the peak 
amplitude is an excellent baryometer.
Upper right: 
in each panel $\oo$ takes on the values indicated.
Notice that the slope at low $\ell$ and the peak amplitude
do not have a simple monotonic dependence on $\oo$. 
Lower left:
in each panel $\ol$ takes on the values indicated.
The largest values of $\ol$ have the largest Doppler peaks. 
As $h$ increases, the peak amplitude decreases 
and the peak position shifts to larger scales.
All models are Harrison Zel'dovich ($n = 1$)  
normalized to the COBE 4-year results.
Figures from Lineweaver \etal 1997a, 1997b.}
}
}
}
\begin{document}
\title{{\bf Gold in the Doppler Hills}:\\
Cosmological Parameters in the Microwave Background
}
\author{C. H. Lineweaver}
\affil{Observatoire Astronomique de Strasbourg\\
11 Rue de l'Universit\'e\\
67000 Strasbourg, France\\
charley@astro.u-strasbg.fr}

\begin{abstract}
Research on the cosmic microwave background (CMB) is
progressing rapidly. New experimental groups are popping up
and two new satellites will be launched.
The current enthusiasm to measure fluctuations in the CMB power 
spectrum at angular scales between $0^{\circ}\!.1$ and $1^{\circ}$ 
is largely motivated by the expectation that CMB determinations 
of cosmological parameters will be of unprecedented precision: cosmological
gold.
In this article I will try to answer the following questions:\\
$\bullet$ What is the CMB?\\
$\bullet$ What are cosmological parameters? \\
$\bullet$ What is the CMB power spectrum?\\
$\bullet$ What are all those bumps in the power spectrum?\\
$\bullet$ What are the current CMB constraints on cosmological parameters?\\
\end{abstract}
\keywords{cosmology, cosmic microwave background, Doppler hills}
\section{What is the Cosmic Microwave Background?}
\label{sec:cmb}

Thirty years ago Penzias and Wilson (1965) discovered excess noise
in their horn antenna in Holmdel, New Jersey. The measured temperature
of this noise was $\sim 3$ K and it did not vary in intensity over
the sky; it was isotropic. They received the Nobel Prize for this
serendipitous discovery of the cosmic microwave background (CMB) radiation.
The prediction of the existence of a CMB and of its temperature 
(Alpher \& Herman 1948) followed by its detection, provides
possibly the strongest evidence for the Big Bang.

\begin{figure}[htbp]     
\centerline{\psfig{figure=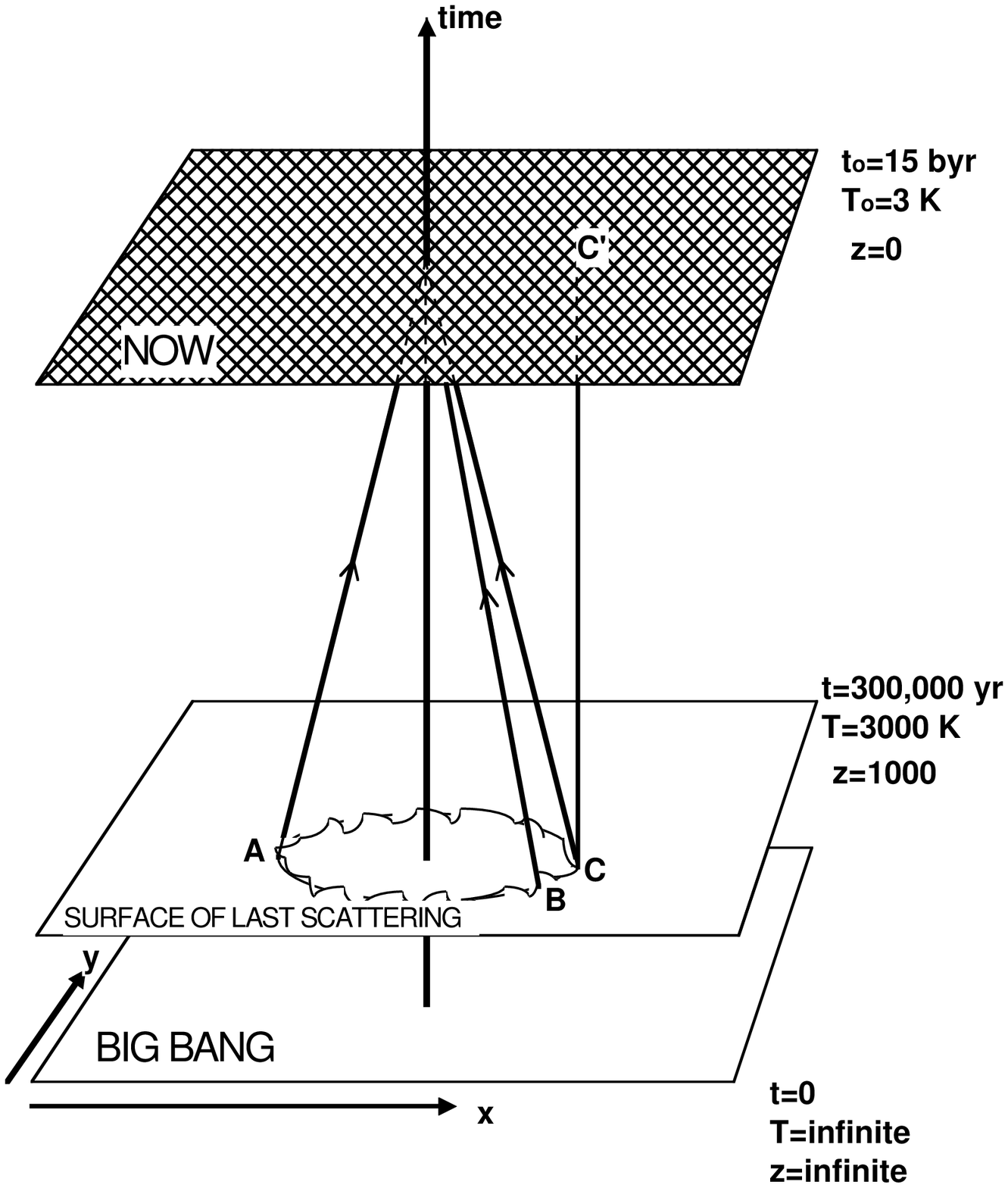,height=9.2cm,width=12cm,bbllx=80pt,bblly=230pt,bburx=550pt,bbury=700pt}}
\caption[Comoving Space-Time and the Surface of Last Scattering]
{{\footnotesize {\bf Comoving Space-Time and the Surface of Last Scattering}. 
The time axis is the world line of the stationary observer who is currently 
located at the apex of the light cone. CMB photons travel from the
wavy circle in the surface of last scattering along the surface of the light
cone to the observer. Points A and C are on opposite sides
of the sky. If the angle between B and C is greater than a few degrees then
B and C have not been in (post-inflationary) causal contact.
The unevenness of the circle represents potential fluctuations at the
surface of  last scattering. The bottom two planes are at fixed times while
the ``NOW" plane moves upward. As it does, the size of the visible Universe
(diameter of the wavy circle) increases. The object seen at C is currently at C'.
We are beginning to determine cosmological parameters by measuring the angular
fluctuations in the temperature of the photons from the circle on the 
surface of last scattering. 
The surface of last scattering, or cosmic photosphere, is at a redshift
of $z \approx 1000$ and is the boundary between the present cool transparent 
universe and the hot opaque Universe of the past.  
CMB photons are valuable fossils which have been studied by dozens of groups 
in efforts to more precisely determine their spectrum
and spatial fluctuations.
Since the most remote quasars are at $z \approx 5$, CMB structures at 
a redshift $z \approx 1000$ are also the most distant objects ever observed. 
Although very distant today, at the time they emitted the light, they were only
$\sim 6\;h^{-1}\:$ Mpc away from us, closer than the Virgo cluster today.
Figure from Lineweaver 1994.}}
\label{fig:spacetime2}
\end{figure}

The observable Universe is expanding and cooling. Therefore in the past
it was hotter and smaller. The CMB is the after-glow of thermal 
radiation left over from this hot early epoch. It is the redshifted 
relic of the Big Bang. 
The CMB is a bath of photons coming from every direction with 
wavelengths about as big as these letters.
There are about $415$ of them in every cubic centimeter of the Universe. 
These are the oldest photons one can observe (see Figure \ref{fig:spacetime2}).
Their long journey towards us has lasted 99.997\% of the age of the 
Universe; a journey which began when the photons were last scattered 
by free electrons of the ubiquitous cosmic plasma, when the Universe 
was $1000$ times smaller and the temperature $1000$ times 
higher than the CMB is today.
The CMB contains 

\noindent
information about the Universe at redshifts much 
larger than the redshifts of galaxies or quasars.
It is a unique tool for probing the early Universe.

To a very good approximation the CMB is a flat featureless blackbody; 
there are no anisotropies; the temperature is a constant
in every direction ($T_{o}=2.728 \pm 0.004\:$K$\:(95\%$CL) Fixsen \etal 1996).
This near isotropy was the reason it took more than 25 years to detect
anisotropies in the CMB.
However the galaxies around us are clustered on scales from 1 Mpc (our
Local Group) up to $\sim~100$ Mpc (great walls, sheets and voids).
If these structures formed from overdensities
which gravitationally collapsed, the overdensities must
have been present in the early universe and must have produced temperature
anisotropies in the CMB.
In the Spring of 1992, the COBE DMR team announced 
the discovery of anisotropies in the CMB (Smoot \etal 1992).
Since then the field of CMB-cosmology has blossomed.

\section{What are Cosmological Parameters?}
\subsection{Friedman Robertson Walker (FRW) Universe}

Cosmological parameters are the important ingredients 
of any cosmological model. If we work within General Relativity and 
add the hypothesis that the Universe is homogeneous and isotropic then the
Einstein equations reduce to the Friedmann equation with its relatively 
few parameters:

\be
\label{eq:friedmann}
\left(\frac{\dot{a}}{a}\right)^{2} = H_{o}^{2}\left[ 
\frac{\Omega_{o,rel}}{a^{4}} +
\frac{\Omega_{o,non-rel}}{a^{3}} +
\frac{\Omega_{o,curv}}{a^{2}} +
\frac{\Omega_{\Lambda}}{a^{0}} 
\right]
\ee

\begin{itemize}
\item $\mathbf {a}$: the scale factor used to parametrize the global expansion (or shrinking)
of the Universe. In the Big Crunch (or looking backwards towards the Big Bang)
$a \rightarrow 0$.
\item
$\mathbf {H_{o}}$: the present value of Hubble's parameter. In terms of the scale 
factor 
$H_{o} = (\dot{a}/a)_{o}$.
The units are $km/s/Mpc$ and it is sometimes written 
dimensionlessly as $h =H_{o}/(100\: km/s/Mpc)$.
The subscript ``${}_{o}$'' refers to the present time.
\item
$\mathbf {\Omega_{o,i}}$: the current dimensionless density of component $i$ of the 
Universe expressed in units of the critical density 
($\rho_{crit}= 3H_{o}^{2}/8\pi G$).
Thus for example the physical density of baryons is 
$\rho_{B} = \ob\; \rho_{crit}$ and the measurement of $\rho_{B}$
gives limits on $\ob h^{2}$.
The critical density marks the boundary between eternally expanding universes
and recollapsing universes. 
\item
$\mathbf {\Omega_{rel}}$: the density of relativistic matter, e.g., hot dark matter (HDM),
neutrinos and radiation for which $pc > m_{o}c^{2}$. 
$\Omega_{rel} = \oh + \on + \og$.
\item
$\mathbf {\Omega_{non-rel}}$: the density of non-relativistic matter, e.g., 
cold dark matter (CDM) or baryons, $\Omega_{non-rel} = \oc + \ob$.
\item
$\mathbf {\Omega_{o}}$: the total matter density of the Universe,
$\Omega_{o}= \Omega_{o,rel} + \Omega_{o,non-rel}$.
\item
$\mathbf {\Omega_{curve}}$: the curvature density of the Universe.
$\Omega_{o,curve}= -kc^{2}/H_{o}^{2}$. 
The factor $k \in [+1, 0, -1]$ for the cases of closed, flat or
open geometries respectively, corresponding to 
$\oo + \ol = [\:<1,\:1,\:>1]$ respectively. 
\item
$\mathbf {\Omega_{\Lambda}}$: vacuum density of the Universe.
$\Omega_{\Lambda} = \frac{\rho_{\Lambda}}{\rho_{crit}}=
\frac{\Lambda c^{2}}{3H_{o}^{2}}= \frac{\Lambda c^{2}}{3 H_{o}^{2}}$. 
The cosmological constant $\Lambda=\frac{8\pi G}{c^{2}} \rho_{\Lambda}$.
\end{itemize}
Evaluating the Friedmann equation at the present ($a=a_{o}=1$) 
provides the constraint $\oo + \Omega_{o, curve} + \ol = 1$.
Inflation implies that the Universe is flat which means
$\Omega_{curve}=0$ and $\oo + \ol= 1$.
In the standard CDM model (flat universe, no cosmological constant, 
$\oo = 1 \approx \Omega_{o,non-rel} >> \Omega_{o,rel}$),
equation (\ref{eq:friedmann}) can be integrated to yield the age of the Universe
$t_{o}= 6.52\; h^{-1}$ Gyr. For more generic cases, $t_{o}= f(h,\oo,\ol)$ 
(see e.g., Kolb \& Turner 1990).
\subsection{Perturbed FRW}

An FRW universe is perfectly isotropic and homogeneous;
a boring universe without galaxies, stars or 
dense perturbations like ourselves. So we need to add perturbations 
to the model. We parametrize these perturbations in terms of the 
normalization and slope of the CMB power spectrum.
There are two families of perturbations which influence the CMB power 
spectrum: scalar (= density) perturbations and tensor 
(= gravitational wave) perturbations,
and correspondingly there are 2 normalizations and slopes.
\begin{itemize}
\item 
$\mathbf {Q}$: the quadrupole normalization of the power spectrum,
usually expressed in $\mu$K.
$Q$ was determined accurately by the COBE satellite.
If $Q=0$ there are no perturbations. 
Inflation does not predict this normalization.
The tensor analog is $\mathbf {T}$.

\item
$\mathbf {n}$: the slope of the primordial power spectrum, observable
today at the largest angles. The tensor analog is $\mathbf {n_{T}}$.
\end{itemize}


\subsection{Summary}

The perfectly homogeneous and isotropic FRW model is parametrized by  
\be
\mathbf {h,\: \oo,\: \ol,\:\Omega_{curve}.}
\ee
The total mass density $\oo$ has several components which are 
also parameters: 
\bea
\mathbf {\oo}&=& \mathbf {[\Omega_{o,non-rel}] +  [\Omega_{o,rel}]}\\
             &=& \mathbf{ [\ob + \oc] + [\oh + \on + \og]}.
\eea
Perturbations to the FRW universe are parametrized by 
\be
\mathbf {n, n_{T}, Q, T}.
\ee
Cosmological parameters are important because they tell us
\begin{itemize}
\item the ultimate destiny of the Universe: 
$f(\oo, \ol)$
\item the age and size of the Universe: 
$f(h, \oo, \ol)$ 
\item the composition of the Universe: $f(\ob, \oc, \oh, \on, \og)$
\item the origin of structure
\end{itemize}
What makes these parameters even more important and what
makes CMB-cosmology such a hot subject is that
in the near future measurements of the CMB angular power spectrum
will determine these parameters with the unprecedented precision of a few
\% (Jungman \etal 1996).
%
\begin{figure}[bth]     
\centerline{\psfig{figure=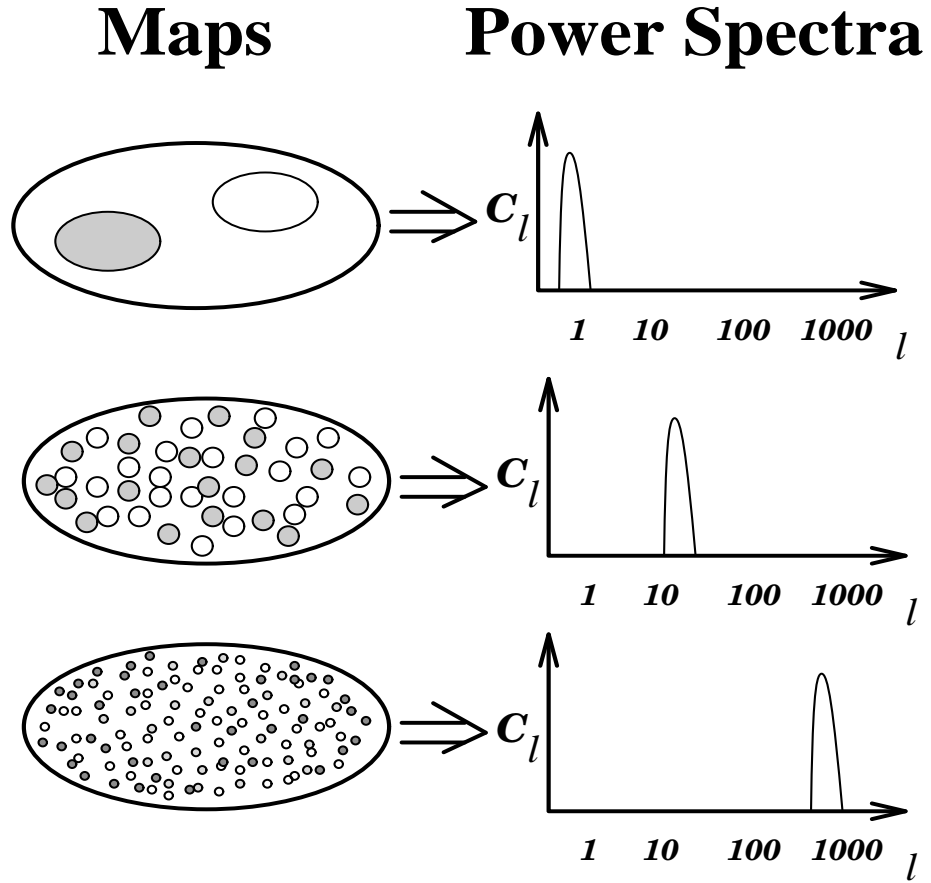,height=12cm,width=12cm}}
\caption[mapcl]
{{\footnotesize {\bf Simple Maps and their Power Spectra}.
If a full-sky CMB map has only a dipole (top), it's power spectrum 
is a delta function at $\ell = 1$ . If a map has only temperature 
fluctuations on an angular scale of $\sim 7^{\circ}$ (middle) then 
all of the power is at $\ell \sim 10$. If all the hot and cold spots 
are even smaller (bottom) then the power is at high $\ell$.
}}
\label{fig:mapcl}
\end{figure}

\section{What is the CMB power spectrum?}

Similar to the way sines and cosines are used in Fourier decompositions
of arbitrary functions on flat space, spherical harmonics can be used
to make decompositions of arbitrary functions on the sphere.
Thus the CMB temperature maps 
are conveniently written as:
\be
\Delta T(\theta, \phi) = \sum_{\ell, m} a_{\ell m} Y_{\ell m}(\theta, \phi).
\ee
The power spectrum is the sum of the squares of the coefficients
\be
C_{\ell} = \frac{1}{2\ell + 1} \sum_{m} a_{\ell m}^{2}.
\ee
See Figure \ref{fig:mapcl} for a brief $C_{\ell}$ initiation.
If the matter power spectrum is written in scaleless form
as $P(k) = A\:k^{n}$, then the radiation power spectrum at scales
larger than a few degrees ($\ell \lsim 20$)
becomes
(Bond \& Efstathiou 1987)

\be
C_{\ell} =  Q^{2} \frac{4\pi}{5} 
\frac{\Gamma(\ell + \frac{n-1}{2})}{\Gamma(\ell + \frac{5-n}{2})}
\frac{\Gamma(\frac{9-n}{2})}{\Gamma(\frac{3+n}{2})},
\ee
where $n$ is the slope of the power spectrum,
$Q^{2}$ is the normalizing quadrupole amplitude (analogous to $A$
and is just another way of writing $C_{2}$)
and $\ell$ sets the angular scale (analogous to the linear scale $k$).
If $n=1$ (as implied by inflation, consistent with the COBE measurements
and first proposed by Harrison (1970) and Zel'dovich (1972)), then 
\be
C_{\ell} =\frac{24\pi}{5}\frac{Q^{2}}{\ell(\ell+1)}
\ee
thus
\be
\ell(\ell + 1)C_{\ell} = \mbox{constant}.
\ee
This is why the y-axis of CMB angular power spectra are labeled with
some function of $\ell(\ell + 1)C_{\ell}$ and why the plotted spectra are
flat for $\ell \lsim 20$ (see Figure \ref{fig:data}).


\subsection{Horizons and Angular Scales}
\label{sec:horizons}
\begin{figure}[h]     
\centerline{\psfig{figure=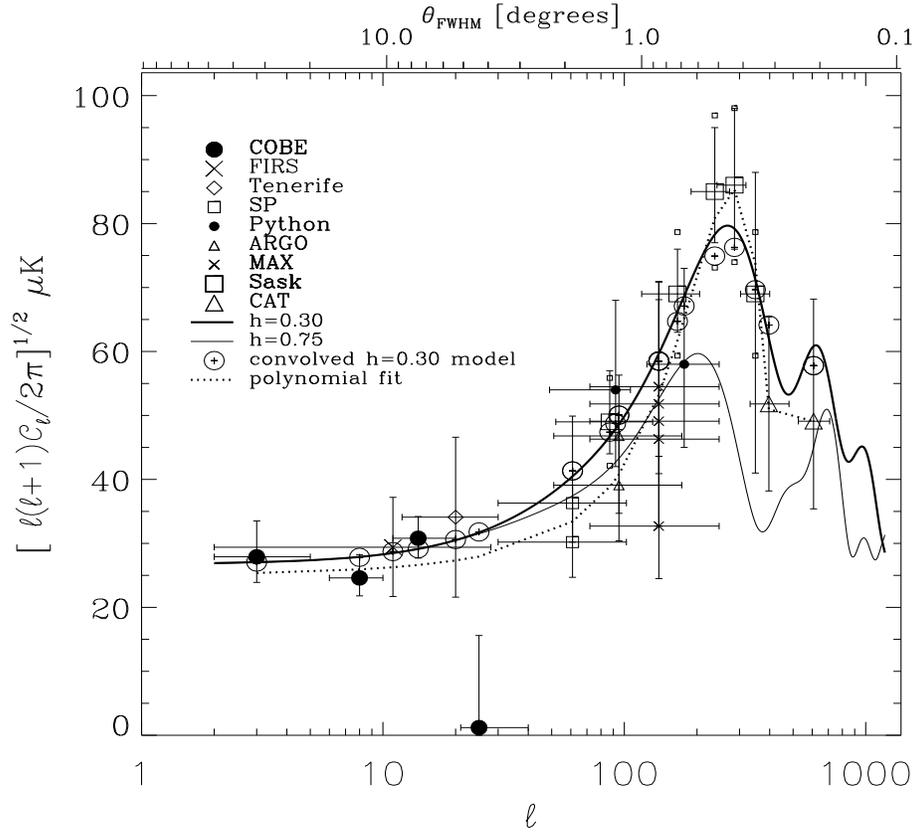,height=11cm,width=12cm}}
\caption[data]
{{\footnotesize {\bf CMB Data}.
A compilation of 24 of the most recent
measurements of the CMB angular power spectrum spanning the region
$2 \lsim \ell \lsim 600$. COBE measurements are at the largest scales
while the CAT interferometer measurements are at the smallest.
Models with $h=0.30$ and $h=0.75$ are superimposed
(both are $\oo = 1$, $\ob=0.05$,  $n=1$, $Q=18\;\mu$K).
The low-$h$ value is preferred by the data (see Figure \ref{fig:hob3}).
The dotted line is a 5th order polynomial fit to the data.
Two satellites,  MAP and Planck Surveyor, are expected to yield precise 
spectra for $\theta_{FWHM} \gtrsim 0^{\circ}\!.3$ ($\ell \lsim 400$) and 
$\theta_{FWHM} \gtrsim 0^{\circ}\!.2$ ($\ell \lsim 700$) respectively.
The angular scale is marked at the top.
Figure from Lineweaver \etal (1997a).
}}.
\label{fig:data}
\end{figure}

\begin{figure}[h]     
\centerline{\psfig{figure=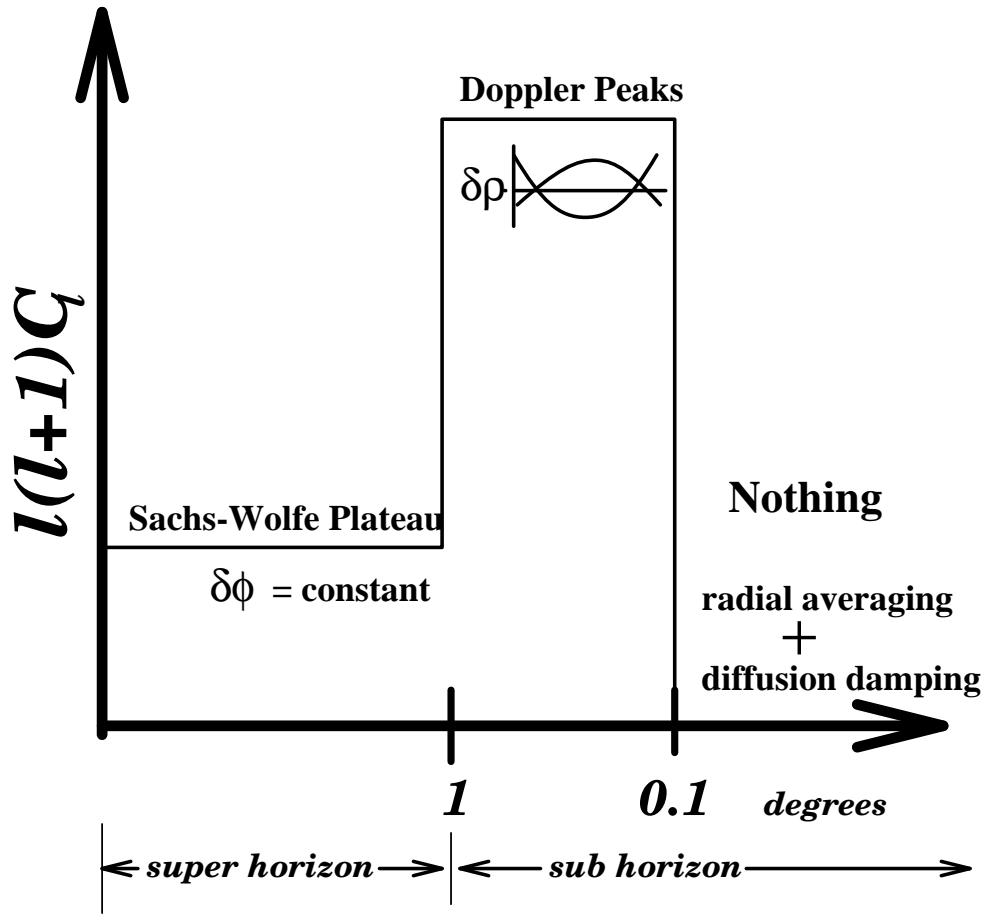,height=12cm,width=12cm}}
\caption[simplewaynesworld]
{{\footnotesize {\bf Simplified CMB Power Spectrum}.
The CMB power spectrum can be crudely divided into three regions.
The Sachs-Wolfe Plateau caused by the scale independence of 
gravitational potential fluctuations which dominate the spectrum
at large super-horizon scales.
The horizon is the angular scale corresponding to $c t_{dec}$ where
c is the speed of light and $t_{dec}$ is the age of the Universe at
decoupling. The Doppler peaks on scales slightly smaller than the 
horizon are due to resonant acoustic oscillations analogous to mellifluous 
bathroom singing (see Figure \protect\ref{fig:acoustic2}).
At smaller scales there is nothing because the finite thickness of the
surface of last scattering averages
small scale fluctuations along the line of sight. Diffusion damping
(photons diffusing out of small scale fluctuations) also suppresses
power on these scales.
}}.
\label{fig:firstapprox}
\end{figure}

To get a rough understanding of the power spectra in Figure 
\ref{fig:data}, we can divide up the plot into super-horizon and sub-horizon
regions as is done in Figure \ref{fig:firstapprox}.
The angular scale corresponding to the particle horizon
size is the boundary between 
super- and sub-horizon scales.
The size of a causally connected region on the surface of last scattering
is important because it determines the size over which astrophysical processes
can occur.
Normal physical processes can act coherently only over sizes smaller
than the particle horizon and could not have produced 
the structure in the COBE maps and {\it a fortiori} could not have produced
the better than one part in $10^{4}$ homogeneity of the entire CMB sky.

A causally connected Hubble patch at last scattering
subtends an angular size (for an observer today) of
\be
\theta_{H} \approx  1^{\circ}\:\Omega_{o}^{1/2}
\left(\frac{z_{dec}}{1000}\right )^{-1/2}.
\ee
%

\begin{figure}[h]     
\centerline{\psfig{figure=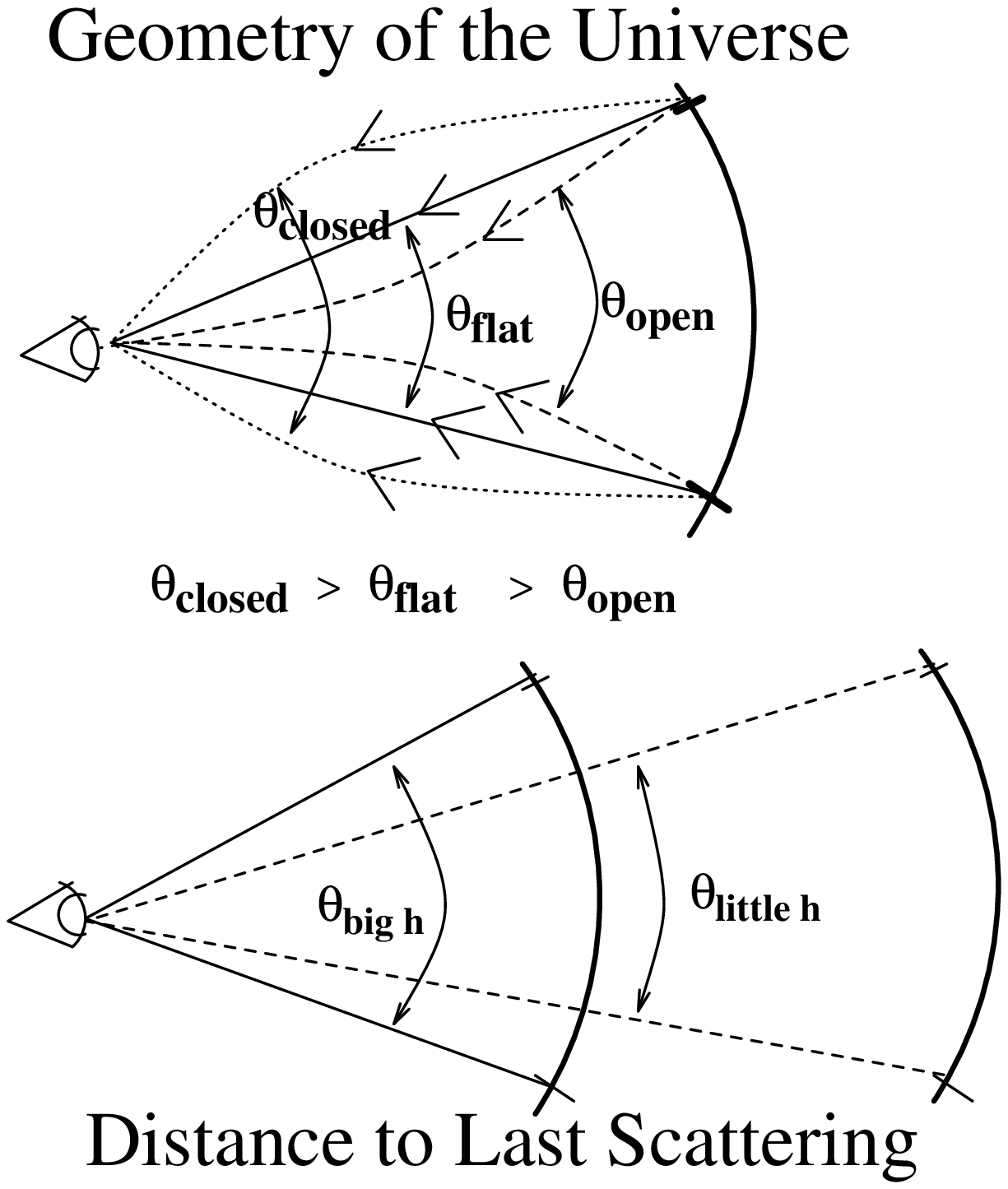,height=13cm,width=12cm}}
\caption[geometry]
{{\footnotesize {\bf Effects of Geometry and Distance}.
Observers are on the left. The surface of last scattering is the
thick curved line on the right.
In the top panel, the same physical scale (on the right) subtends 
different angular scales depending on the geometry.
In closed universes (geometry of the surface of a sphere,
 $\oo + \ol > 1$, k = +1),  the angle is largest.
In open universes (geometry of the surface of a saddle,
 $\oo + \ol < 1$, k = -1),  the angle is smallest.
In flat universes (Euclidean geometry of a plane,
 $\oo + \ol = 1$, k = 0),  the angle is between the other two.
In the flat universe of the bottom figure, the angle subtended by a 
given physical scale depends on the distance to the surface of
last scattering (see Section \protect\ref{sec:horizons}).
Summary: all the features of the power spectrum are shifted to smaller scales
in open universes and for small $h$.
}}.
\label{fig:geometry}
\end{figure}

\clearpage
\noindent
Thus as $\oo \uparrow$, $\theta_{H} \uparrow$ and as 
$z_{dec} \uparrow$, $\theta_{H} \downarrow$ (see Figure \ref{fig:geometry}).
The angle $\theta$ subtended by an object of size $ct$ at an angular 
distance $d_{ang}$ is $\theta \sim ct/d_{ang}$. Thus the angular scale
associated with the peak of the power spectrum is
\be
\label{eq:lpeak}
\ell_{peak} \sim \frac{1}{\theta_{peak}} 
            \sim \frac{d_{ang}(h, \Omega_{o}, \Omega_{\Lambda})}{c\;t_{dec}}
            \sim \frac{h^{-1}f(\Omega_{o},\Omega_{\Lambda})}{(\Omega_{o} h^{2})^{-1/2}}
            \sim \frac{f(\Omega_{\Lambda})}{h},
\ee
where $\theta_{peak}$ is the angular scale of the Doppler peak.
The physical scale of the peak oscillations 
is some fixed fraction of the horizon $\propto c\;t_{dec}$.
The time of decoupling scales as $(\Omega_{o} h^{2})^{-1/2}$ 
and the angular distance $d_{ang}= h^{-1} f(\Omega_{o},\ol)$.
In flat models $\Omega_{o} = 1-\ol$ and $f(\Omega_{o}, \ol) \rightarrow f(\ol)$
where 
\be
f(\ol) = \int_{0}^{z_{dec}} 
\frac{dz}{\left[ (1+z)^{3}- \ol(1+z)^{3}\right]^{1/2}}.
\ee
Inserting this into equation (\ref{eq:lpeak}) yields the 
monotonic relation: when $\ol \uparrow$, $\ell_{peak}\uparrow$ (see Figure 5b).
Equation (\ref{eq:lpeak}) says that
when $\frac{f(\ol)}{h}\uparrow \ell_{peak} \uparrow$.
The $h$ scaling can be understood as the effect of larger universes:
a given physical size at a larger distance subtends a smaller angle
(see Figures \ref{fig:geometry} and 5b).

There is structure in the DMR maps on super-horizon scales.
How did it get there? Inflation
is invoked to explain this apparently acausal structure 
(see the contribution by Liddle in this volume).
If defect models of structure formation are correct then this acausality
is only apparent; low $z$ defects produced the large scale anisotropies.
If inflation is correct, the apparent causal disconnection
of the spots in the DMR maps means we are looking much further back than
the epoch of last scattering.
The structure that one sees in the DMR maps may represent a glimpse of 
quantum fluctuations at the inflationary epoch $\sim 10^{-32}$ seconds 
after the Big Bang, showing us scales $\sim 10^{16}$ times smaller than the 
atomic structure seen with the best ground-based microscopes.
For more on the DMR instrument as a microscope see
Lineweaver (1995).

\section{What are all those Bumps in the Power Spectrum}
\subsection{Decoupling and the Surface of Last Scattering}

At about 300,000 years after the bang, the
Universe had cooled down enough to allow the free electrons and protons
to combine to form neutral hydrogen. This period is known as decoupling. 
This neutralization of the plasma
allowed photons to free stream in all directions. 
Before decoupling the Universe was an opaque fog of 
free electrons, afterwards it was transparent. 
The boundary is called decoupling, recombination, the cosmic photosphere 
or the surface of last scattering;
the surface where the CMB photons were Thomson scattered for the last time before
arriving in our detectors.

Decoupling occurs when the CMB temperature has dropped to the point when
there are no longer enough high energy photons in the 
CMB to keep hydrogen ionized; $\gamma + H \leftrightarrow e^{-} + p$.
Although the ionization potential of hydrogen is 13.6 eV ($T \sim 10^{5}$ K)
decoupling occurs at $T \approx 3000$ K.
The high photon to proton ratio ($\eta \approx 10^{9}$) allows the high
energy tail of the Planck distribution to 

%
\clearpage
\noindent
\putboxedtopfigs

\noindent
\putboxedbottomfigs

\clearpage
\noindent
keep the comparatively small number
of hydrogen atoms ionized until this much lower temperature
(for details check out the Saha equation).

The temperature of the CMB as a function of redshift is $T(z)=T_{o}(1+z)$.
Decoupling occurs at a fixed temperature $T(z_{dec}) = constant$.
As the Universe cools down $T_{o} \downarrow$, thus
the surface of last scattering recedes from us with
an ever-increasing redshift, $z_{dec} \propto \frac{1}{T_{o}(t)}$.

%

\subsection{Anisotropy Mechanisms in a Perturbed Robertson-Walker Universe}
\label{sec:animech}

The temperature of CMB photons can be influenced by any field which couples to
photons. There are three:
\begin{itemize}
\item Gravity $\phi(\vec{r})$, by gravitational red and blue shifts
\item Density  $\rho(\vec{r})$, by compression heating and rarefaction cooling
\item Velocity $v(\vec{r})$, by scattering from moving charged particles 
(Doppler effect).
\end{itemize}
The dominant effects on the CMB produced by these fields
occur at the surface of last scattering, i.e., at a distance 
$|\vec{r}| \approx 6000\; h^{-1}$ Mpc from the observer in the direction 
of the line of sight (Figure \ref{fig:lss}).
The differential temperature of the CMB in direction $\vec{r}$, 
$\delta T(\vec{r})=T(\vec{r})-\T$, can be expressed as
a function of the potential $\phi$, the density fluctuations $\delta$
and the velocity $\vec{v}$.
\be
\label{eq:full}
\frac{\delta T(\vec{r})}{T_{o}}= 
\frac{\phi(\vec{r})}{c^{2}} + \frac{1}{3}\delta(\vec{r}) - \vec{r} \cdot \frac{\vec{v}(\vec{r})}{c}
+ \frac{2}{c^{2}}\int \dot{\phi}(\vec{r},t)\: dt,
\ee
or in words,
\be
{\mbox Temperature} = {\mbox Gravity} + {\mbox Density} +{\mbox Velocity}
+{\mbox Changing}\:{\mbox Gravity.}
\ee
Notice that all four terms in equation (\ref{eq:full})
are independent of the frequency of the
radiation. This spectral flatness is used by observers
to distinguish CMB anisotropies from Galactic and extragalactic foregrounds.
In the next two pages I will try to explain how these different terms
influence the CMB on different scales.

\subsection{Large Super-horizon Scales}
$\bullet$ Gravity

On angular scales larger than a few degrees, the cold and hot spots in the 
CMB maps are caused by the red- and blue-shifting of photons leaving
primordial gravitational potential fluctuations (Sachs \& Wolfe 1967).
That is, photons at the surface of last scattering loose energy  climbing 
out of potential valleys and gain energy falling down potential hills;
and these valleys and hills have different amplitudes as a function of position on the sky. 
Hills produce hot spots while valleys produce cold spots.
The Pound-Rebka experiment used the Mossbauer effect and confirmed the 
existence of a gravitational redshift of magnitude $\phi/c^{2}$,
the first term of equation (\ref{eq:full}).

\clearpage
\begin{figure}[h,p,t]     
\centerline{\psfig{figure=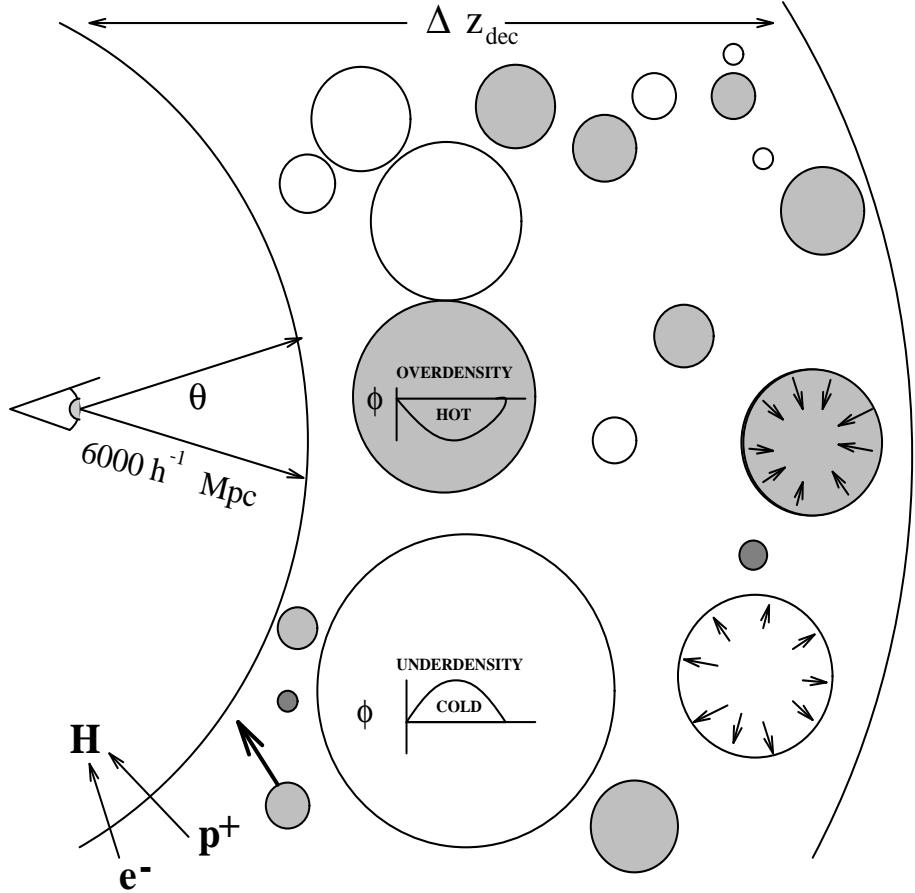,height=12cm,width=12cm}}
\caption
{{\footnotesize {\bf Fluctuation Production at Last Scattering}.
Gravity ($\phi$), density ($\rho$) and 
velocity $(v)$ fields couple to the CMB photons and produce 
temperature anisotropies at the surface of last scattering.
The grey circles are hot potential wells and the white circles are
cool potential hills.
Adiabatic conditions specify that the locations of the CDM potential
wells coincide with the positions of the baryon-photon overdensities, 
and potential hills coincide with the baryon-photon underdensities.
In climbing out of the potential wells, the initially hot
photons become gravitationally redshifted and end up cooler than average.
Similarly, in falling down the potential hills, the initially cooler
photons become hotter than average. Thus on the largest scales
(where these two effects dominate) the cool spots in the COBE maps
are regions of overdensity.
Bulk velocities of the plasma
are indicated by the arrow on the grey spot in the lower
left but are not expected on super horizon scales.  
On horizon scales, matter is 
falling into potential wells and falling down potential hills
analogous to the bulk flow velocities measured in local 
pre-virialized objects.
Acoustic velocities are indicated by the radial arrows in the 
well and hill on the right (see Figure \protect\ref{fig:acoustic2}).
}}.
\label{fig:lss}
\end{figure}

\clearpage
\noindent
$\bullet$ Density

The initial conditions are usually selected to be 
adiabatic and less commonly isocurvature.
With adiabatic initial conditions the locations of the overdensities 
in the baryon-photon fluid coincide with the locations of the potential wells.  This 
leads to a partial cancelling of the gravity and density terms 
(see Figure \ref{fig:lss}). 
On super-horizon scales $\delta \approx -2\phi/c^{2}$, thus
the sum of the gravity and density terms is 
$\phi/3c^{2}$ (gravity wins).

With isocurvature initial conditions the curvature from CDM potential
wells is compensated by coinciding underdensities of the baryon-photon
fluid. No curvature (= ``isocurvature'') is the result.
The gravity and density terms do not cancel, in fact they add coherently
leading to relatively more power on super-horizon scales
compared to the adiabatic case.

\noindent
$\bullet$ Velocity

The  $\vec{r} \cdot \vec{v}(\vec{r})/c$ term is the standard Doppler effect 
applied to radiation.
The velocity can be conveniently decomposed
\be
\vec{v}(\vec{r})=\vec{V}_{\odot} + \vec{v}_{dec}(\vec{r})
\ee
where
$\vec{V}_{\odot}$ is the velocity of the observer,
i.e., the velocity of the Sun with respect to the CMB 
and $\vec{v}_{dec}$ is the velocity of the last scattering plasma with respect 
to the CMB.
The Doppler term from $\vec{V}_{\odot}$ produces the large observed 
dipole, known also as the ``Great Cosine in the Sky''.
The measurement of this dipole tells us how fast we are moving with
respect to the rest frame of the CMB 
(Lineweaver \etal 1995, Lineweaver \etal 1996).
When we make a CMB map and remove the mean,
the next largest feature visible at 1000 times smaller amplitude
is the dipole. But the amplitude of this kinetic  dipole is 
$\sim 100$ {\bf larger} than the anisotropies of the CMB power
spectrum.

On large scales equation (\ref{eq:full}) becomes
\be
\label{eq:fullsw}
\frac{\delta T(\vec{r})}{T_{o}}= 
\frac{  \phi(\vec{r})       }{3c^{2}}
- \frac{  \vec{r} \cdot \vec{V}_{\odot}   }{c}.
\ee
When we remove the dipole from the maps we are left with only
the combined gravity/density term of the Sachs-Wolfe effect, $\phi/3c^{2}$.

On super-horizon scales, the physical decomposition we are making here is
ambiguous, i.e., gauge dependent. One gauge's adiabatic compression is 
another gauge's gravitational redshift, but the observed 
$\delta T(\vec{r})/T_{o}$ is gauge independent (see e.g. Hu 1995).

\subsection{Small Sub-horizon  Scales}
$\bullet$ Gravity

\begin{figure}[h,p,t]     
\centerline{\psfig{figure=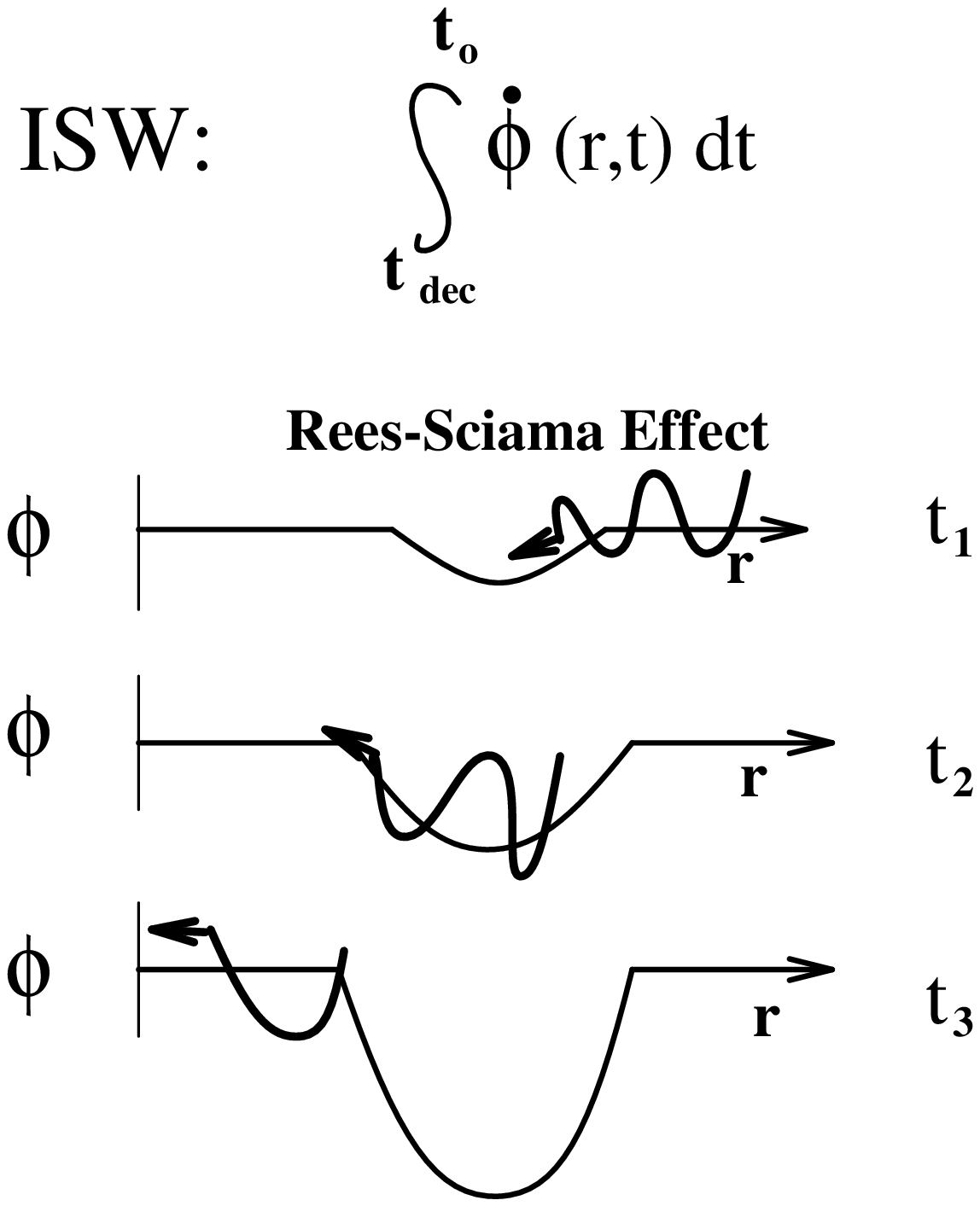}}
\caption
{{\footnotesize {\bf Integrated Sachs-Wolfe (ISW) Effect}.
Consider an overdensity that is growing such as a collapsing 
proto-cluster of galaxies.
CMB photons cross such structures on their way to us.
Falling into a shallow potential well and then climbing out of
it when it is deeper results in a net redshift of the photons.
This is known as the Rees-Sciama effect and is a specific case of the
more generic integrated Sachs-Wolfe (ISW) effect.
}}.
\label{fig:isw}
\end{figure}

The integrated Sachs-Wolfe effect is gravitational redshifting 
when the CMB photons fall into shallow
potential valleys and climb out of deep valleys (Figure \ref{fig:isw}). 

The early ISW effect is due to the self-gravity of the 
photons just after $z_{dec}$. 
Since photon potentials do not grow with the same scaling as the 
non-relativistic matter, $\dot{\phi}\neq 0$.
Near decoupling, $\Omega_{\gamma}(z \sim z_{dec})$ is non-negligible and
if we let $h \rightarrow  0.20$, $z_{eq} \rightarrow z_{dec}$,
which means that $\og$ is larger at decoupling and that
the contribution from the Early ISW effect increases. 

\clearpage
\clearpage
\noindent
The late ISW effect is also from  $\dot{\phi}\neq 0$ and is
produced in non-flat universes ($k\neq 0$) or when $\Lambda \neq 0$.
It is ``late'' in the sense that in the limit as  $a\rightarrow \infty$ (late times)
the last two terms in the Friedmann equation control 
the expansion
(see Tegmark 1995 and Hu 1995 for more details).

\begin{figure}[h,p,t]     
\centerline{\psfig{figure=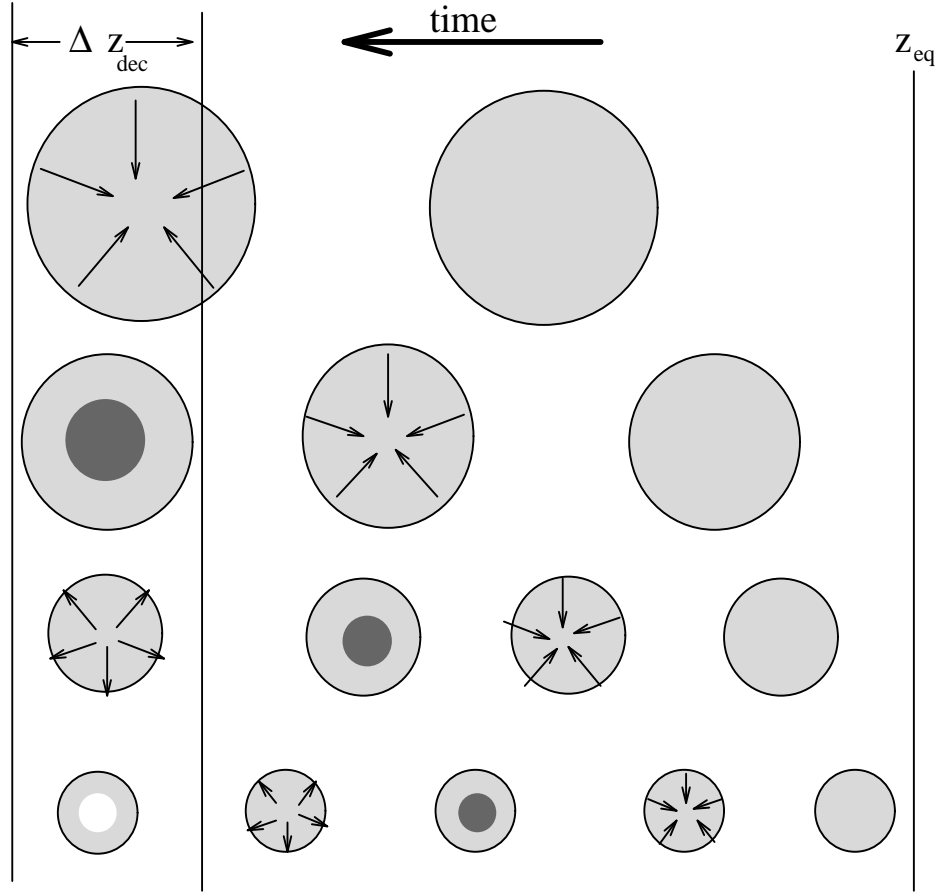,height=12cm,width=12cm}}
\caption
{{\footnotesize {\bf Seeing Acoustic Oscillations}.
The grey spots are CDM potential wells of four different sizes evolving in time.
The arrows represent velocities of the baryon-photon fluid.
After $z_{eq}$, in the matter dominated era,
the baryon-photon fluid can begin to collapse into potential 
wells which enter the horizon. Acoustic oscillations
on scales smaller than the sound horizon can begin to oscillate.
The imprint of these acoustic oscillations is left in the CMB photons
when the Universe becomes transparent during the period marked
``$\Delta z_{dec}$''.
Thus we see acoustic oscillations in the snap-shot of the
Universe called the surface of last scattering.
The top row corresponds to the largest scale Doppler contribution and
contributes power at scales slightly larger than the main acoustic peak
(few degrees).
It is caught at decoupling with maximum velocity.
The second row corresponds to the main acoustic peak in the power 
spectrum at an angular scale of $\sim 0.5^{\circ}$ 
(see Figure \protect\ref{fig:data}).
This is inappropriately called the ``Doppler peak''. It is caught 
at maximum compression (= hot) when the velocities are minimal.
Potential hills of the same size (not shown here) produce a
rarefaction peak ( = cold).
The third row is the second Doppler peak which fills in the 
first valley of the power spectrum ($\sim 0.3^{\circ}$). 
The last row is the second acoustic peak ($\sim 0.2^{\circ}$). 
It is a rarefaction peak (= white spot) for
potential wells and a compression peak for potential 
hills (not shown here). 
The compression/rarefaction
peaks are 90$^{\circ}$ out of phase with the velocity peaks.
}}.
\label{fig:acoustic2}
\end{figure}

After matter-radiation equality, the growth of CMB potential
wells and hills drives acoustic oscillations (see Figure \ref{fig:acoustic2}).

\noindent
$\bullet$ Density

The correlated combination of density and velocity
fluctuations are acoustic waves.
Since we are dealing essentially with a single baryon-photon fluid,
(the electrons  couple the baryons tightly to the photons) adiabatic
compression and rarefaction of this fluid creates hot and cold spots
that can be seen (Figure \ref{fig:acoustic2}).

\noindent
$\bullet$ Velocity

Plasma at the surface of last scattering
can have a velocity due to bulk motions or to
acoustic oscillations which are 90 degrees out of phase with density 
fluctuations.
Figure \ref{fig:acoustic2} displays these acoustic oscillations at
different scales.


For simplicity, in equation (\ref{eq:full}), we have
assumed a Robertson-Walker metric and therefore do not consider
differential expansion as a source of anisotropy.
Additionally, we do not include the Vishniac (1987) 
and Sunyaev-Zel'dovich (1972) effects. These post-decoupling
effects contribute to small angular scale anisotropies.
We also do not include the more speculative anisotropies due to
topological defects (monopoles, strings, walls, textures)
or any contribution from a possible rotation 
of the Universe (Barrow, Juszkiewicz \& Sonoda 1985).
We also do not include polarization anisotropies.
For excellent reviews of this subject 
and more details see Tegmark (1994), Hu (1995), Bunn (1997)  and 
Hu, Sugiyama \& Silk (1997).

\section{What are the current CMB constraints on cosmological parameters?}
\label{sec:results}

The current enthusiam to measure fluctuations in the CMB power 
spectrum at angular scales between $0^{\circ}\!.1$ and $1^{\circ}$ is largely motivated by the 
expectation that CMB determinations of cosmological parameters will be of 
unprecedented precision.
In such circumstances it is important to estimate and keep track of what we 
can already say 
about the cosmological parameters. 
In two recent papers (Lineweaver \etal 1997a \& 1997b)
we have compiled the most recent CMB measurements, used a fast Boltzmann 
code to calculate model power spectra (Seljak \& Zaldarriaga 1996) 
and, with a $\chi^{2}$ analysis, we have compared the data to the power spectra
from several large regions of parameter space.

In Lineweaver \etal (1997a) we considered
COBE-normalized flat universes with $n=1$ power spectra. We used predominantly
goodness-of-fit statistics to locate the regions of the $h - \Omega_{b}$
and $h - \Lambda$ planes preferred by the data.
In Lineweaver \etal (1997b) we obtained $\chi^{2}$ values over the 4-dimensional parameter
space $\chi^{2}(h,\Omega_{b}, n, Q)$ for $\Omega = 1$, $\Lambda = 0$ models.
Projecting and slicing this 4-D matrix gives us the error bars around the minimum $\chi^{2}$
values.
Here we summarize several of our most important results.

One of the difficulties in this analysis is the 14\% absolute calibration uncertainty of
the 5 important Saskatoon points which span the dominant adiabatic peak in the spectrum
(Figure \ref{fig:data}).
We treat this uncertainty by doing the analysis 
\clearpage

\clearpage
\noindent
three times: all 5 points at their
nominal values (`Sk0'), with a 14\%
increase (`Sk+14') and a 14\% decrease (`Sk-14').
Sk+14 and Sk-14 are indicated by the small squares in Figure \ref{fig:data} above and below the nominal
Saskatoon points.
Leitch \etal (1997) report a preliminary relative calibration of Jupiter and CAS A 
implying that the Saskatoon calibration should be $-1\% \pm 4\%$.
Reasonable $\chi^{2}$ fits are obtained for Sk0 and Sk-14.

\begin{figure}[h,p,t]     
\centerline{\psfig{figure=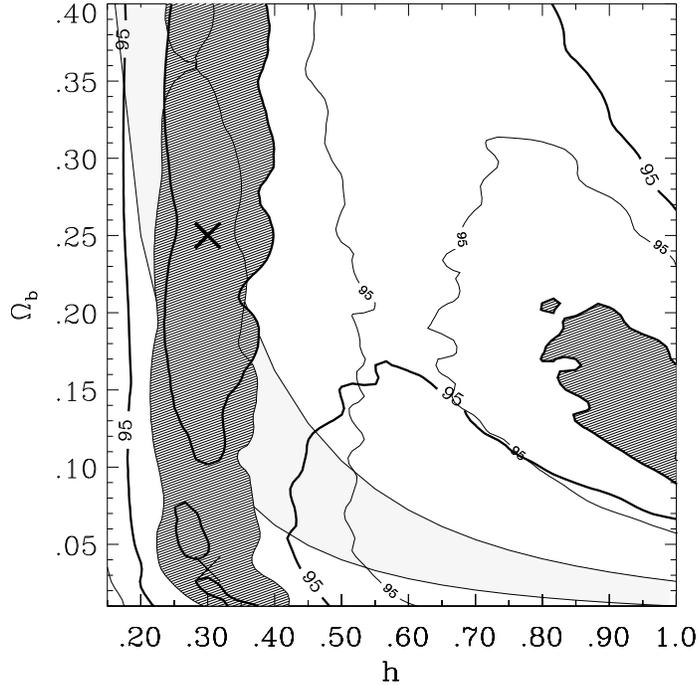,height=10cm}}
\caption[hob3]
{{\footnotesize 
Constraints on Hubble's Constant
The dark grey areas denote the regions of parameter space favored by the
CMB data. They are defined by $\chi^{2}_{min} + 1$ for 
Sk0 and Sk-14 (minima marked with thick and thin `x' respectively)
(cf. Section \protect\ref{sec:results}).
`95' denotes the $\chi^{2}_{min} + 4$ contours for
Sk0 (thick) and Sk-14 (thin).
The light grey band is from Big Bang nucleosynthesis ($ 0.010 < \Omega{b}\:h^{2} < 0.026$). 
The parameters $n$ and $Q$ have been marginalized.
In the $H_{o}$ result quoted, we neglect the region
at $H_{o} \sim 100$ with $\Omega_{b} \sim 0.15$.
This figure shows clearly that lowering the calibration by
14\% {\it {\bf does not}} favor higher values of $H_{o}$. 
Figure from Lineweaver \etal 1997b.
}}
\label{fig:hob3}
\end{figure}

\begin{figure}[htbp]     
\centerline{\psfig{figure=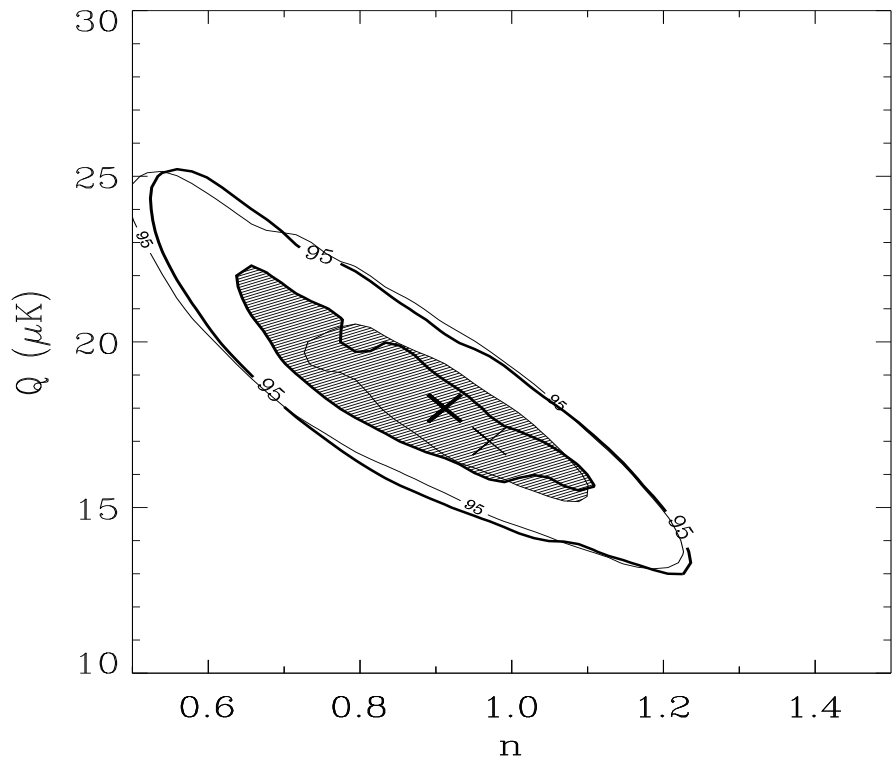,height=11cm,width=12cm}}
\caption[nq3]
{{\footnotesize {\bf Constraints on $n$ and $Q$}.
A precise measurement of the spectral slope $n$ and normalization $Q$
using recent CMB data.
$n=0.93^{+0.17}_{-0.16}$ and
$Q=17.5^{+3.5}_{-2.5}\:\mu$K.
See Lineweaver \etal (1997b) for further details.
}}
\label{fig:nq3}
\end{figure}

In the context of the flat models tested, our $\chi^{2}$ analysis yields:
$H_{o} = 30 ^{+13}_{-9}$ (Figure \ref{fig:hob3}),
$n=0.93^{+0.17}_{-0.16}$ and
$Q=17.5^{+3.5}_{-2.5}\:\mu$K (Figure \ref{fig:nq3}).
The $n$ and $Q$ values are consistent with previous estimates while the 
$H_{o}$ result is surprisingly low.

\clearpage
\noindent
For each result, the other 3 parameters have been marginalized.
This $H_{o}$ result has a negligible dependence on the Saskatoon 
calibration, i.e., lowering the Saskatoon calibration from 0 to -14\% 
does not raise the best-fitting $H_{o}$ in flat models.
The inconsistency between this low $H_{o}$ result and $H_{o} \sim 65$ results
will not easily disappear with a lower Saskatoon calibration. 
Our results are valid for the specific models we considered:
$\Omega=1$, CDM dominated, $\Lambda = 0$, Gaussian adiabatic initial
conditions, no tensor modes, no early reionization,
$T_{o}=2.73$ K, $Y_{He}=0.24$, no defects, no HDM.

There are many other cosmological measurements which are consistent 
with such a low
value for $H_{o}$ (Bartlett \etal 1995, Liddle \etal 1996). 
For example, we calculated a joint likelihood based on the 
observations of galaxy cluster baryonic fraction, Big Bang nucleosynthesis 
and the large scale density fluctuation shape parameter, $\Gamma$. 
We obtained $H_{o}\approx 35^{+6}_{-5}$.

With two new CMB satellites to be launched in the near future 
(MAP $\sim 2001$, Planck Surveyor $\sim 2005$) and half a dozen new 
CMB experiments coming on-line (23 groups), the future looks bright
for CMBers ( see Page 1997).
%

\clearpage
\acknowledgments
I am grateful to my collaborators D. Barbosa, A. Blanchard and J.G. Bartlett.
I want to thank Martin Hendry for his scottish weltanschauuang,
David Valls-Gabaud for his anxious professionalism, 
Khalil Chamcham for his international optimism and Hannah Quaintrell 
for being there.
I am also grateful to Nour-Eddine Najid and Idriss Mansouri of the
Faculty of Sciences, Ain-Chock, Hassan II University, Casablanca 
for helpfully arranging my two public lectures. 
Wayne Hu made several helpful suggestions.
I acknowledge support from NSF/NATO post-doctoral fellowship 9552722.

\begin{question}{Dr. Liddle}
What did you say about $\chi^{2}$?
\end{question}
\begin{answer}{Dr. Lineweaver}
We can talk about that later.
\end{answer}

\begin{question}{Dr. Hermit}
If what you say is true then everybody else is just plain
wrong about the value of Hubble's constant.
\end{question}
\begin{answer}{Dr. Lineweaver}
Hmmm. We'll see.
\end{answer}



\end{document}